\documentclass[twocolumn]{aastex63}
\usepackage{amssymb}
\usepackage{amsmath}
\usepackage{graphicx}

\newcommand{\HII}{\hbox{H\,{\sc ii}}}
\newcommand{\HI}{\hbox{H\,{\sc i}}}

\shortauthors{Mao et al.}

\begin{document}

\title{\textbf{
HOLMBERG\,IX: A UNIQUE, INFANT BUT INACTIVE GALAXY \\
as Revealed via a Multiwavelength Approach} \\
}

	\correspondingauthor{Ye-Wei Mao}
	\email{ywmao@gzhu.edu.cn}
	
\author{Ye-Wei Mao}
\affiliation{Center for Astrophysics, GuangZhou University, GuangZhou 510006, China; \textbf{ywmao@gzhu.edu.cn}}
\affiliation{Department of Astronomy, School of Physics and Materials Science, GuangZhou University, GuangZhou 510006, China}

\author{Luis C. Ho}
\affiliation{Kavli Institute for Astronomy and Astrophysics, Peking University, BeiJing 100871, China}
\affiliation{Department of Astronomy, School of Physics, Peking University, BeiJing 100871, China}

\author{Alexei V. Moiseev}
\affiliation{Special Astrophysical Observatory, Russian Academy of Sciences, Karachai-Cherkessian Republic 369167, Russia}
\affiliation{Sternberg Astronomical Institute, Lomonosov Moscow State University, Universitetsky pr. 13, Moscow 119234, Russia}

\author{Oleg V. Egorov}
\affiliation{Astronomisches Rechen-Institut, Zentrum f\"{u}r Astronomie der Universit\"{a}t Heidelberg, M\"{o}nchhofstra\ss e 12-14, D-69120 Heidelberg, Germany}

\author{Andrej M. Sobolev}
\affiliation{State Key Laboratory of Radio Astronomy and Technology, XinJiang Astronomical Observatory, Chinese Academy of Sciences,
150 Science 1-Street, Urumqi, XinJiang, 830011, China}
\affiliation{Ural Federal University, 19 Mira Street, 620002 Ekaterinburg, Russia}

\begin{abstract}
In this Letter, we report a novel discovery of unique characteristics for the tidal dwarf galaxy (candidate) Holmberg\,IX via a multiwavelength investigation.  New observations are taken for deeply mapping H$\alpha$ emission and combined with archival/published data for comprehensively probing dust, gas, and stellar populations in this galaxy.  We find in Holmberg\,IX a dearth of dust incompatible with its rich gas and metal; globally young stellar populations with prominent far-ultraviolet but deficient and marginal H$\alpha$ emissions, distinct from other tidal dwarf galaxies ever known.  By assuming a normal initial mass function (IMF), Holmberg\,IX is suggested to be born $\sim 130$\,Myr ago from a bursty star formation event, which then rapidly ceased, with very few stars formed in the past $\sim 80$\,Myr that demarcates a lower age limit for the galactic mainbody; current star formation occurs only in outskirts, bringing a conundrum about the reason for the recent quenching in such a gas-rich environment.  Contradicting the general expectation for tidal dwarf galaxies hosting continuous star formation, the present quiescence implies Holmberg\,IX currently staying in a rarely-seen transient period.  Without star formation continuing, Holmberg\,IX is likely transforming into a dwarf spheroidal galaxy, or oppositely into a(n) (ultra-)diffuse system which will probably dissolve in the end.  Instead, if Holmberg\,IX possesses peculiar IMF and hosts low-mass, weak-H$\alpha$ star formation, it is able to maintain long-term survival in its current status.  On whichever evolutionary pathway in reality, Holmberg\,IX appears as a special case updating conventional understandings of tidal dwarf galaxies and hinting potential existence of similar analogs in the Universe.
\end{abstract}

\keywords{galaxies: dwarf - galaxies: individual (Holmberg\,IX) - galaxies: irregular - galaxies: star formation - galaxies: stellar content}

\section{\textbf{INTRODUCTION}}\label{Sec_Intro}

Gravitational interactions between large galaxies are likely to strip materials out of the interacting galaxies due to tidal effect and consequently generate new stellar systems with relatively small size, usually termed as "\emph{tidal dwarf galaxies}" \citep[TDGs,][]{1978IAUS...77..279S, 1992Natur.360..715B, 1992A&A...256L..19M}.  Suffering from turbulence induced by multiple gravitational fields, TDGs are not able to assemble into steady structures and hence appear irregular in morphology.  Dissimilar to dwarf spheroidal/elliptical galaxies lying in a quiescent state, TDGs are expected to continuously form stars from birth to the present time, which yields strong emission lines prominently featuring in their spectra \citep{1999IAUS..186...61D}.

Destinies of TDGs after a cessation of star formation (hereafter denoted as "SF") are unclear and debatable.  There have been two opposite pathways proposed to chart the evolution.  Without SF continuing, some of TDGs are likely to dissolve into diffuse or filamentary systems and perhaps eventually dissipate in the Universe \citep{1995AJ....110..140H, 2010ApJ...717L.143M, 2013MNRAS.436..839S}, whereas some others are suggested with a high probability to condense and transform into long-lived compact spheroidal/elliptical systems \citep{2006A&A...456..481B, 2007MNRAS.376..387M, 2013MNRAS.429.1858D}.  In whatever evolution scenario, TDGs can hardly be preserved in present shapes for long time if stars stop forming, yet there have not been TDGs discovered with no current SF, possibly because such a period is too short to be captured.

The work reported in this Letter targets a specific candidate of TDGs: Holmberg\,IX, also named DDO\,66, UGC\,5336, or PGC\,28757. This is a dwarf irregular galaxy accompanying the spiral galaxy NGC\,3031 (M81), with the distance $\sim 3.6$ Mpc and the size comparable to the Small Magellanic Cloud but much lower luminosity \citep{1969ArA.....5..305H, 1974A&A....32..117B, 2009ApJS..183...67D}. Residing on a neutral hydrogen bridge connecting M81, NGC\,3034 (M82), and NGC\,3077 \citep{1979A&A....75...97V, 1994Natur.372..530Y, 2018ApJ...865...26D}, Holmberg\,IX is considered to originate from tidal debris of a gravitational interaction between the three primary galaxies as "the second generation" in the M81 group \citep{2002A&A...396..473M}.

Stellar populations in Holmberg\,IX have been studied through observations at optical and near-infrared (NIR) broadbands taken with \emph{Hubble Space Telescope} and large ground-based telescopes \citep{2008ApJ...676L.113S, 2008ApJ...689..160W, 2015ApJ...809L...1O}. By jointly analyzing color-magnitude diagrams (CMDs) and stellar density maps, these studies have found that observable stars younger than $\sim 250$ Myr show a high level of spatial concentration; older stars have a relatively uniform distribution and a slightly increasing trend toward M81 in spatial density, comparable to those in other fields at the same distance from M81. This feature suggests the old stellar component unaffiliated with Holmberg\,IX but associated with the outer halo of the adjacent companion M81. Therefore, stars belonging to Holmberg\,IX are suggested not to evolve over 250 Myr. The upper limit $< 250$ Myr on age is in agreement with the time for the gravitational interaction in the M81 group $\sim 200$--$300$\,Myr ago \citep{1999IAUS..186...81Y}, in support of the hypothesis of the tidal origin for Holmberg\,IX.

However, existing optical or NIR broadband photometric diagnostics are not capable of reaching young stellar populations and thereby constraining the lower age limit, which makes current SF state unclear.  Excited by high-energy photons from O-type stars with lifetimes of $\lesssim 10$ Myr, the H$\alpha$ emission line stands in a good position to trace SF currently occurring \citep[][hereafter denoted as K98]{1998ARA&A..36..189K}. There have been some observations of H$\alpha$ emission in Holmberg\,IX taken by telescopes with small diameters, but these studies have yielded considerable deviation in estimates of SF rate \citep[SFR;][]{1994ApJ...427..656M, 2008ApJS..178..247K, 2011SerAJ.183...71A}.

Aimed at exploring the youngest stellar populations and current SF state in Holmberg\,IX, in this work, we conduct new observations of H$\alpha$ emission in Holmberg\,IX via not only direct imaging but also Fabry-Perot interferometer (FPI) scanning, for the purpose of spatially integrated and resolved measurements, respectively.  Archival/published data are also compiled as a complement.  The proximity of Holmberg\,IX brings intrinsically high spatial resolution, making this TDG candidate an ideal object for a spatially resolved study.

In the following content, we describe the H$\alpha$ observations and compilation of the ancillary data in Section \ref{Sec_Data}, present the results of data analysis in Section \ref{Sec_Result}, and discuss the implication of the results in Section \ref{Sec_Disc}.

\begin{deluxetable*}{lrrrcc}
\tabletypesize{\normalsize}
\tablecaption{\textbf{Statistics of the UV, H$\alpha$, and IR Observations and Photometry}}
\tablewidth{0pc}
\tablehead{
\colhead{Waveband} & \colhead{$\lambda_\mathrm{eff}$} & \colhead{$\Delta\lambda$} & \colhead{$t_\mathrm{Exp}$} & \colhead{$Flux_\mathrm{\,Aper}$} & \colhead{$Flux_\mathrm{\,Sele}$} \\
\colhead{(1)} & \colhead{(2)} & \colhead{(3)} & \colhead{(4)} & \colhead{(5)} & \colhead{(6)}
}
\startdata
 \emph{GALEX}~FUV & 1516 & 268 & 14,707 & 2.293~$\pm$~0.106 & 0.699~$\pm$~0.032 \\
 \emph{Swift}-UVOT~UVW2 & 1928 & 657 & 389,667 & 2.678~$\pm$~0.074 & 0.804~$\pm$~0.022 \\
 \emph{GALEX}~NUV & 2267 & 732 & 29,422 & 2.972~$\pm$~0.082 & 0.907~$\pm$~0.025 \\
 \emph{Swift}-UVOT~UVW1 & 2600 & 693 & 158,330 & 3.096~$\pm$~0.086 & 0.918~$\pm$~0.025 \\
 BTA H$\alpha$-image  & 6571 & 13 & 900 & 4.335~$\pm$~0.252 & \nodata \\
 BTA H$\alpha$-FPI & 6571 & 13 & 14,400 & \nodata & 3.907~$\pm$~0.117 \\
 % \emph{WISE}~w4  & 22000 & 4110 & 170 & $< 1.096$ & $< 0.517$ \\
 % \emph{Spitzer}-MIPS~24 & 24000 & 5320 & 220 & $<  1.295$ & $< 0.611$ \\
 % \emph{Herschel}-PACS~70 & 70000 & 2130 & \nodata & $< 70.062$ & $< 33.035$ \\
 % \emph{Herschel}-PACS~160 & 160000 & 6850 & \nodata & $< 53.492$ & $< 25.222$ \\
 % \emph{Herschel}-SPIRE~250 & 250000 & 6730 & \nodata & $< 24.852$ & $< 11.718$ \\
 % \emph{Herschel}-SPIRE~350 & 350000 & 9520 & \nodata & $< 14.167$ & $< 6.680$ \\
 % \emph{Herschel}-SPIRE~500 & 500000 & 1840 & \nodata & $< 6.635$ & $< 3.128$
\enddata
\tablecomments{~
Columns: (1) Name of each filter in order of effective wavelength; (2) Effective wavelength in units of \AA; (3) Bandwidth in units of \AA; (4) Total exposure time in units of seconds; (5) Flux for Holmberg\,IX as a whole obtained with the aperture photometry described in Section \ref{Sec_Data_Phot}, in units of mJy for the UV data, and 10$^{-14}$ erg\,s$^{-1}$\,cm$^{-2}$ for the H$\alpha$ data; (6) Flux for the area of the effective H$\alpha$ emission in Holmberg\,IX obtained with the selective photometry described in Section \ref{Sec_Data_Phot}, in units of mJy for the UV data, and 10$^{-14}$ erg\,s$^{-1}$\,cm$^{-2}$ for the H$\alpha$ data.
}
\label{Tab_Stat}
\end{deluxetable*}

\section{\textbf{DATA AND MEASUREMENT}}\label{Sec_Data}

\subsection{H$\alpha$ Emission-line Images}\label{Sec_Data_Ha}

Observations of Holmberg\,IX with FPI were carried out on December 19--21 in 2019, by use of the 6 m Big Telescope Alt-azimuth (BTA) equipped with a prime focus multimode focal reducer SCORPIO-2 \citep{Afanasiev2011} at the Special Astrophysical Observatory of the Russian Academy of Sciences.  The total exposure time was 14,400\,s for 40 interferograms corresponding to different gaps between  FPI plates, reaching the $3 \sigma$ sensitivity limit $4 \times 10^{-18}$\,erg\,s$^{-1}$\,cm$^{-2}$\,arcsec$^{-2}$.  Data reduction was performed in compliance with \citet{Moiseev2015}, producing a datacube with a wide-field image ($6\arcmin \times 6\arcmin$) multiplying a high-resolution spectrum ($R \approx 16\,000$) at the position of each pixel (0\arcsec.7\,pixel$^{-1}$).  The final spatial resolution was $\sim 2\arcsec.8$ as FWHM of the point-spread function (PSF) convolved to the largest seeing among all of the combined frames.  The H$\alpha$ FPI map was created via Voigt-profile fitting of the emission-line spectra with the signal-to-noise ratio $> 3$, and is shown in Figure \ref{Fig_UVHaIR} .

BTA Observations of Holmberg\,IX in the direct-imaging mode of SCORPIO-2 were performed on December 24 in 2020, with a narrowband filter centered at 6571\,\AA~ with FWHM $\approx 13$\,\AA.  The total exposure time was 900\,s with the spatial resolution $\sim 1\arcsec.4$ (equal to the seeing in the observation night) and the sampling scale 0\arcsec.4\,pixel$^{-1}$.  A continuum image was observed with a filter having a similar bandwidth centered near H$\alpha$ and used to create a net emission-line image.  Images of spectrophotometric standard stars, taken before and after the observations of the galaxy during the same night, were used to calibrate counts into absolute energetic flux.  This H$\alpha$ narrowband image, also shown in Figure \ref{Fig_UVHaIR}, was applied to flux calibration of the H$\alpha$ FPI map.

\subsection{UV Archival Images}\label{Sec_Data_UV}

Holmberg\,IX was imaged at ultraviolet (UV) bands with deep exposure by the \emph{Galaxy Evolution Explorer} \citep[\emph{GALEX};][]{2011ApJS..192....6L} and the \emph{Swift} UV/Optical Telescope \citep[\emph{Swift}-UVOT;][]{2011AJ....141..205H}. In our study, four-band UV images were adopted, among which far-UV (FUV), near-UV (NUV) were downloaded from Guest Investigators database of \emph{GALEX} at the Multimission Archive at Space Telescope Science Institute (MAST) website,\footnote{\url{http://galex.stsci.edu/}} and UVW2, UVW1 were retrieved from the \emph{Swift}-UVOT survey at the \emph{SkyView} virtual observatory.\footnote{\url{https://skyview.gsfc.nasa.gov/}}  Basic information on the UV and H$\alpha$ observations including effective wavelength, bandwidth, and exposure time is listed in Table \ref{Tab_Stat}.

\subsection{Image Processing}\label{Sec_Data_Proc}

Global background in the \emph{GALEX} and \emph{Swift}-UVOT images was estimated by polynomial fitting along both spatial axes in source-masked images with $\sim 1^\circ \times 1^\circ$ fields, and then subtracted from the original images \citep[the background fitting method is addressed in][and relevant articles therein]{2014ApJ...789...76M}.  In order to carry out spatially resolved measurements, unification of PSFs is a requisite. In this work, the BTA, \emph{GALEX}, and \emph{Swift}-UVOT images were convolved to match the \emph{GALEX} NUV PSF (FWHM $\sim 5\arcsec.05$, the lowest resolution amongst the images), with adapted kernels offered in \citet{2011PASP..123.1218A}.  The PSF-matched images were registered on the same scale (1\arcsec.5\,pixel$^{-1}$) and the same coordinate by using the SWarp software \citep{2002ASPC..281..228B}.

Since Holmberg\,IX is shrouded by the extended halo of M81 \citep{2008ApJ...676L.113S, 2015ApJ...809L...1O}, it is necessary to remove contribution of M81 starlight from the \emph{GALEX} and \emph{Swift}-UVOT images.  The removal of the local contamination followed the same process with the large-field background subtraction, but was implemented in a small window ($7 \arcmin \times 7 \arcmin$) with Holmberg\,IX placed at the center.  For the BTA H$\alpha$ emission-line images, the M81 starlight had been removed naturally through continuum subtraction.

In the final step, the images were corrected for Galactic foreground extinction by adopting $A_\mathrm{V}^{\mathrm{Gal}} = 0.22$\,mag quoted from the \citet{2011ApJ...737..103S} extinction map, the \citet{1999PASP..111...63F} extinction curve, and the total-to-selective extinction ratio $R_\mathrm{V} = 3.1$.

Figure \ref{Fig_UVHaIR} is an eight-stamp atlas of Holmberg\,IX imaged through the UV and H$\alpha$ channels, at both of the original and convolved spatial resolutions.  All measurements and analyses presented in below sections are based on the convolved images.

\subsection{Photometry}\label{Sec_Data_Phot}

\begin{figure*}[!h]
\centering
\vspace*{-10cm}
\hspace*{-18mm}
\includegraphics[width=1.18\columnwidth]{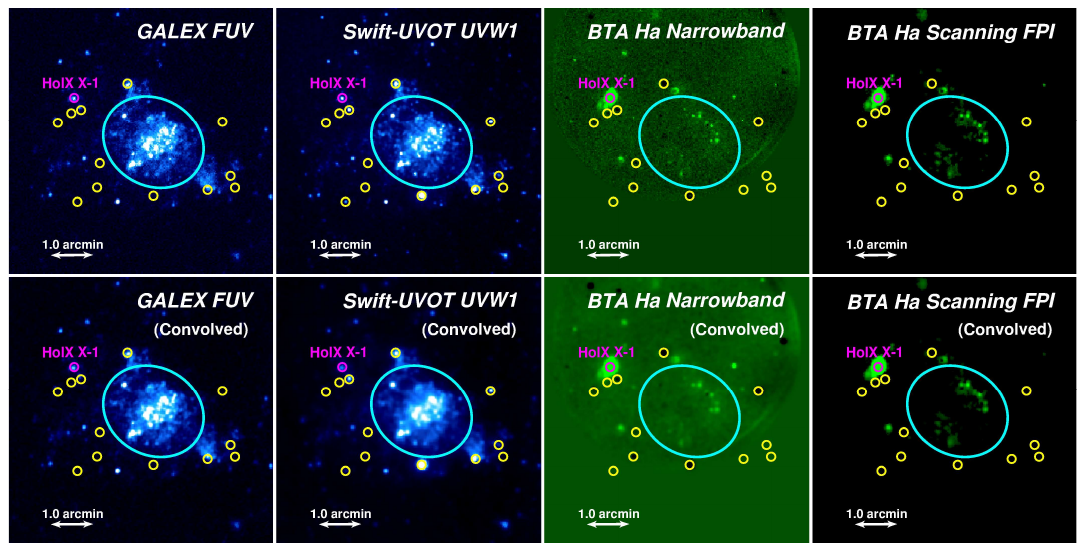}
\\[-9cm]
\caption{Holmberg\,IX imaged at \emph{GALEX} FUV, \emph{Swift}-UVOT UVW1, BTA H$\alpha$ narrowband, and BTA Scanning FPI H$\alpha$ channel (from left to right).  The images are displayed with their original PSF (top row) and after convolution into the common \emph{GALEX} NUV PSF (bottom row).  In each of the images, cyan ellipse represents the photometric aperture for the main body of the galaxy as a whole (described in Section \ref{Sec_Data_Phot}); yellow and magenta circles mark locations of foreground stars and the position of HoIX\,X-1, respectively; the scale ruler at the bottom-left corner in each panel indicates the 1$\arcmin$.0 length; north is up, and east is to the left.} \label{Fig_UVHaIR}
\end{figure*}

\begin{figure*}[!h]
\centering
\includegraphics[width=0.495\columnwidth]{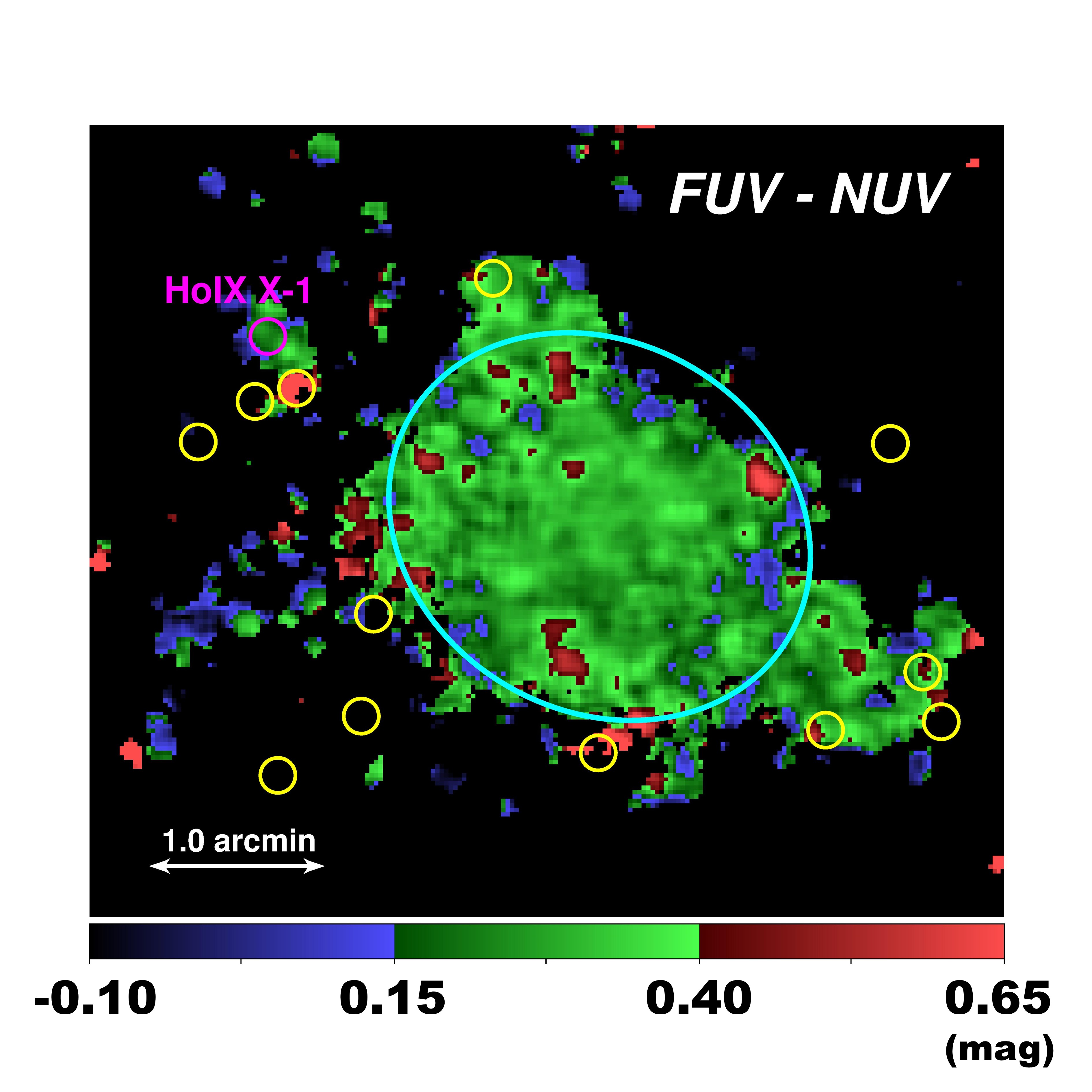}
\vspace*{-1.5cm}
\includegraphics[width=0.495\columnwidth]{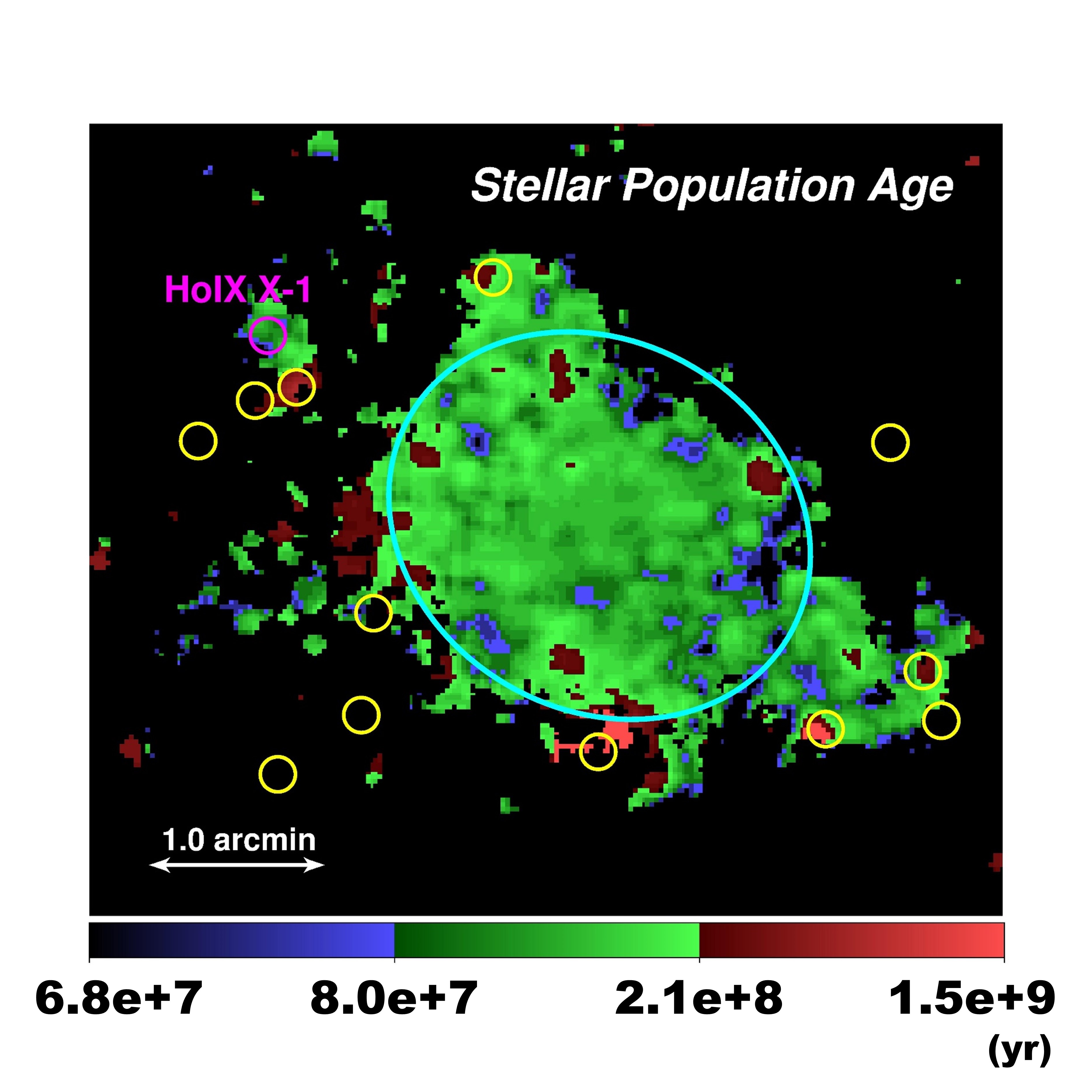}
\hspace*{-1.75cm}
\includegraphics[width=0.61\columnwidth]{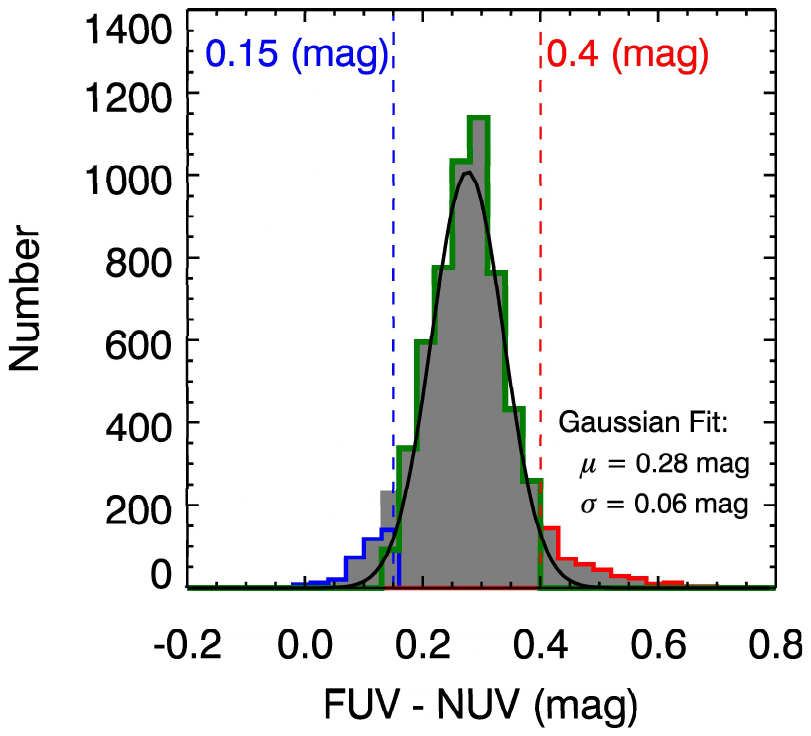}
\hspace*{-2.5cm}
\includegraphics[width=0.61\columnwidth]{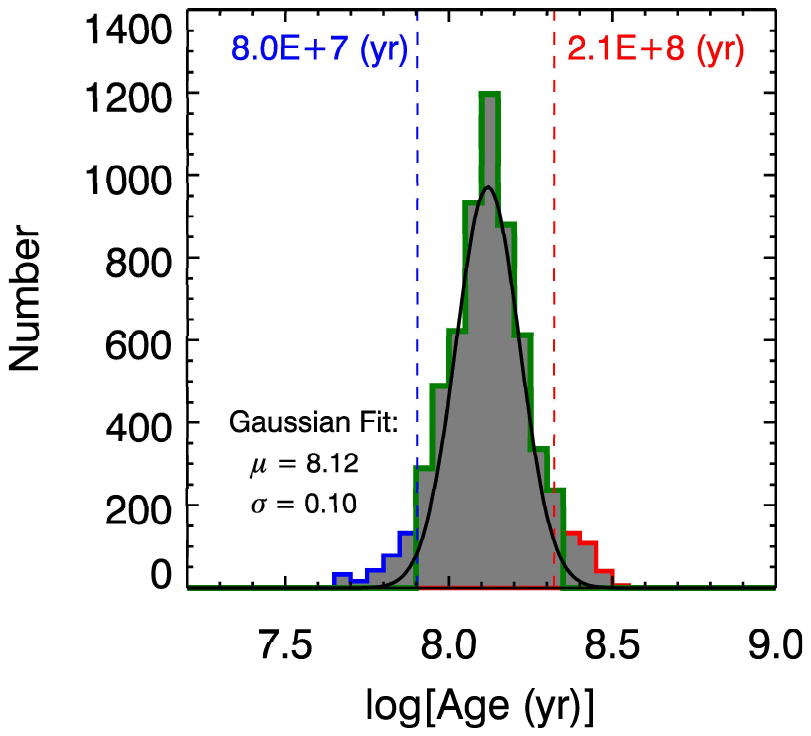}
\hspace*{-1.5cm}
\\[-6cm]
\caption{Top: 2D maps of the color index $\mathrm{FUV}-\mathrm{NUV}$ (left) and stellar population age (right) for Holmberg\,IX, where cyan ellipse, yellow and magenta circles are the same as assigned in Figure \ref{Fig_UVHaIR}; the scale numbers of the color bars are in units of magnitude (left) and year (right), respectively; the scale ruler in the bottom-left corner indicates the 1$\arcmin$.0 length; north is up, and east is to the left.  Bottom: Histograms of the color index $\mathrm{FUV}-\mathrm{NUV}$ (left) and stellar population age (right) for Holmberg\,IX enclosed in the elliptical aperture, derived from corresponding 2D maps in the top panels, where black solid lines are the best-fit Gaussian profiles, and dashed lines demarcate boundaries in accordance with the $2 \sigma$ range, identical to the color bars in the above 2D maps.}\label{Fig_UVcolor}
\end{figure*}

For the purpose of obtaining global properties of Holmberg\,IX, \emph{aperture photometry} was utilized to measure the main body of the galaxy as a whole, by employing an elliptical aperture centered at R.A. = $09^\mathrm{h}57^\mathrm{m}31^\mathrm{s}.95$, Decl. = $69^{\circ}02\arcmin45\arcsec.60$, with the diameters $2\arcmin.5 \times 2\arcmin.1$ (corresponding to the physical scale 2.6\,kpc $\times$ 2.2\,kpc) and the position angle $60^{\circ}$.  The shape of the aperture was determined manually in order to enclose the bulk of galactic brightness but avoid involving the low-brightness edge of the galaxy contaminated by background sources and foreground stars.\footnote{Foreground stars in this sky field, marked with yellow circles in Figures \ref{Fig_UVHaIR}--\ref{Fig_aperUV-UV}, are identified in optical and NIR images obtained via the Beijing-Arizona-Taipei-Connecticut multicolor wide-field survey \citep[i.e., BATC;][]{1996AJ....112..628F, 2005ApJ...630L.133S}, the Sloan Digital Sky Survey \citep[i.e., SDSS;][]{2000AJ....120.1579Y, 2015ApJS..219...12A}, and the Two Micron All Sky Survey \citep[i.e., 2MASS;][]{2000AJ....119.2498J, 2003AJ....125..525J}.}  This photometric aperture is displayed as the cyan ellipse in Figures \ref{Fig_UVHaIR}1, the top panels of Figure \ref{Fig_UVcolor}, and the left panel of Figure \ref{Fig_aperUV-UV}.

The H$\alpha$ FPI map prescribes a domain of H$\alpha$ emission in the galaxy.  In order to specifically measure the H$\alpha$-emission area, an extra type of photometry was conducted in this work to complement the aperture photometry.  This special measurement is defined as "\emph{selective photometry}", by precisely selecting pixels at identical positions of the H$\alpha$ emission inside the elliptical aperture. The area measured with the selective photometry accounts for $\sim 22\%$ of that with the aperture photometry.

Fluxes obtained with the aperture and selective photometry at the UV and H$\alpha$ bands are listed in Table \ref{Tab_Stat}.  It should be noted that the most dominant part of H$\alpha$ luminosity for Holmberg\,IX is produced by the nebula surrounding the ultra-luminous X-ray source HoIX\,X-1 \citep{2008RMxAA..44..301A} and not counted in this work.  The fluxes in Table \ref{Tab_Stat} represent the galactic mainbody enclosed by the photometric aperture rather than completely cover total emissions associated with Holmberg\,IX.  As can be seen from Figure \ref{Fig_UVHaIR}, there are also some outlying sources outside the aperture at each waveband.

In addition to the photometry on a galactic scale, subregions of the galaxy on a smaller scale were also measured in this work.  The subregions were sampled in the \emph{GALEX} FUV image and categorized into two classes.  Emission peaks above $5 \sigma$ level in the FUV image were detected with the SExtractor software \citep{1996A&AS..117..393B} and defined as \emph{FUV clusters}, representing the youngest stellar population inside Holmberg\,IX.  All the other subregions than the FUV clusters were defined as \emph{diffuse regions} and presumed to have evolved into older stages.

Photometry of these subregions was performed in the \emph{GALEX} FUV, NUV, and \emph{Swift}-UVOT UVW2, UVW1 images, by employing circular apertures with the radius $4\arcsec.5$ (corresponding to the local length 80\,pc).  For the diffuse regions, the apertures were placed to cover as much area as possible between the FUV clusters, but at the same time to prevent overlap between each other.  The diffuse regions with AB magnitude $<$ 22\,mag (corresponding to the surface brightness $<$ 26.5\,mag\,arcsec$^{-2}$) at the FUV band were only sampled.  The final sample consists of 63 FUV clusters and 45 diffuse regions.  The placement of the photometric apertures for the galactic subregions is displayed in the left panel of Figure \ref{Fig_aperUV-UV}. % AB magnitudes at the 4 UV bands obtained through the photometry are supplied in Appendix \ref{App}.

Uncertainties assigned to the photometry were estimated as a quadratic sum of background deviation and calibration uncertainties in the relevant images.  The background deviation is calculated as standard deviation at the non-signal area in each background-subtracted image. The calibration uncertainties taken into the quadrature are 0.05 and 0.03\,mag for \emph{GALEX} FUV and NUV, respectively \citep{2007ApJS..173..682M}, 0.03 mag for \emph{Swift}-UVOT bands \citep{2011AIPC.1358..373B}, 3\% for both of the SCORPIO-2 modes of the BTA observations.

\subsection{Stellar Population Synthesis Modeling}\label{Sec_Data_Mod}

Applied to characterizing observational properties, theoretical spectra were extracted from the GALAXEV spectral library of stellar population synthesis \citep{2003MNRAS.344.1000B}.  The parametric space of the spectral model contains six values of metallicity (i.e., $Z$ = 0.0001, 0.0004, 0.004, 0.008, 0.02, and 0.05), the \citet{2003PASP..115..763C} initial mass function (IMF), and the star formation history (SFH) of an instantaneous burst.  The assumed SFH is justified in Appendix \ref{Sec_App} by comparing with more complex forms of SFH.  No dust attenuation was imposed on the spectral model (since Holmberg\,IX appears with no detectable interstellar dust, as presented in the below section).  Monochromatic fluxes in the model were obtained by convolving the theoretical spectra with transmission curves of corresponding filters.

\section{\textbf{ANALYSIS AND RESULTS}}\label{Sec_Result}

\subsection{Dustless Trait Disclosed by IR Inspection}\label{Sec_Result_IR}

The sky field of Holmberg\,IX has been covered by a number of IR observations including the Enhanced Resolution Galaxy Atlas project of the \emph{Wide-field Infrared Survey Explorer} \citep[\emph{WISE};][]{2013AJ....145....6J}, the Local Volume Legacy (LVL) survey of the \emph{Spitzer Space Telescope} \citep[\emph{Spitzer};][]{2009ApJ...703..517D}, and the Very Nearby Galaxies Survey of the \emph{Herschel Space Observatory} \citep[\emph{Herschel};][]{2010A&A...518L..65B}.  We visually inspect the images at \emph{WISE} 22\,$\mu$m, \emph{Spitzer}-MIPS 24\,$\mu$m, \emph{Herschel}-PACS 70, 160\,$\mu$m, and \emph{Herschel}-SPIRE 250, 350, 500\,$\mu$m bands, and find no clustering signal but only random fluctuation throughout the Holmberg\,IX field.  Through tentative photometry in each of the IR images, we find the difference between the average values for the pixels inside and outside the elliptical aperture smaller than the background deviation.  This visual plus tentatively photometric test reflects very trivial dust content (if any) in Holmberg\,IX with the total amount of either hot or cold dust below detection limits of the IR telescopes, coincident with previous studies of this galaxy concluding undetectable IR emission \citep{2017ApJ...837...90D} and negligible dust attenuation \citep{2011AJ....141..205H}.\footnote{\citet{2017ApJ...837...90D} failed to obtain integrated fluxes for Holmberg\,IX above detection limits at 14 IR bands from 5.8 to 850\,$\mu$m; \citet{2011AJ....141..205H} attempted to quantify internal dust attenuation of Holmberg\,IX without IR data, but resulted in an estimate less than the error in measurement.}  On a larger scale in this sky field of the IR images, detectable IR signals are supposed to originate from foreground Galactic cirrus belonging to the Milky Way \citep{2010MNRAS.409..102D} or background residues unrelated with local sources.  In order to corroborate the peculiarly dustless trait for Holmberg\,IX from a more definitive perspective, we perform a systematic analysis by referencing relevant previous studies, which is presented below.

\citet[][hereafter denoted as D23]{2023AJ....165..260D} have conducted a statistical study of spectral energy distributions (SEDs) for 258 nearby galaxies surveyed in the LVL program \citep{2009ApJS..183...67D}, with Holmberg\,IX contained in the sample.  Mid- and far-IR images for LVL galaxies have been obtained with \emph{Spitzer}-MIPS observations and measured with large apertures.\footnote{Although D23 provides IR luminosities for Holmberg\,IX observed with \emph{Spitzer}-MIPS, the extracted photometry is likely to contain large contributions from other sources out of the galaxy, such as foreground Galactic cirrus or background sources as explained above (Nonetheless, this is merely our suspicion without solid evidence at this moment; studies based on deeper and more resolved observations are required for drawing robust conclusions).  In this case, we consider the D23 data as upper limits on the integrated IR luminosities to assess maximal dust content for Holmberg\,IX.}  In D23, dust mass for each of the galaxies has been obtained through SED fits with thoughtful modeling of dust emission, where $M_\mathrm{dust} = 10^{4.53}\,M_{\bigodot}$ for Holmberg\,IX, corresponding to the mass surface density $\Sigma_\mathrm{dust} = 10^{3.50}\,M_{\bigodot}\,\mathrm{kpc}^{-2}$.

The \HI~mass surface density for Holmberg\,IX, as provided in the catalog of \citet[][via observations with the 100 m radio telescope at Effelsberg]{2004AJ....127.2031K}, is $\Sigma_\mathrm{\HI} = 10^{7.90}\,M_{\bigodot}\,\mathrm{kpc}^{-2}$.  CO emission is undetectable for Holmberg\,IX via observations with the James Clerk Maxwell Telescope \citep{2012MNRAS.424.3050W}, which enables assuming no molecular gas in this galaxy.\footnote{Possibility of CO-dark molecular gas existing cannot be ruled out, but absence of molecular gas is in better agreement with the trivial dust content.}  By combining the dust and HI mass densities, we obtain the extremely low dust-to-gas ratio $\sim 4.0\times10^{-5}$, particularly compared with metallicity reference expounded in below paragraphs.

By quoting spectroscopic observations with the Gemini North telescope in \citet{2009ApJ...705..723C}, \citet{2010ApJS..190..233M} have derived the average oxygen abundance $12+\log(O/H) = 8.98 \pm 0.05$ ($\sim 1.95\,Z_{\bigodot}$) for two \HII~regions in Holmberg\,IX, from the strong-line index $R23$ with the \citet[][hereafter denoted as KK04]{2004ApJ...617..240K} theoretical calibration on the basis of photoionization model.  This metellicity corresponds to the dust-to-gas ratio $\sim 0.01$ according to the dust-metal relation \citep{2018ARA&A..56..673G, 2024ARA&A..62..113H}.  By taking the above \HI-gas content into account, the expected surface density of dust mass is $\Sigma_\mathrm{dust} \sim 10^{6}\,M_{\bigodot}\,\mathrm{kpc}^{-2}$, much higher than the above SED-fit estimate.\footnote{Discrepancy in different metallicity diagnostics is a ubiquitous problem.  From the same $R23$, the average oxygen abundance is derived alternatively as $12+\log(O/H) = 8.86 \pm 0.20$ ($\sim 1.48\,Z_{\bigodot}$) with the \citet{1991ApJ...380..140M} theoretical calibration but $8.14 \pm 0.11$ ($\sim 0.28\,Z_{\bigodot}$) with the \citet{2005ApJ...631..231P} empirical calibration.  The former is nearly the same with the KK04 estimate within the error, whereas the latter is much lower and leads to the dust-to-gas ratio $\sim 10^{-4}$--$10^{-3}$, but the far smaller dust amount than the metallicity reference for Holmberg\,IX remains conclusive in this case.}

It is intriguing that the metallicity for Holmberg\,IX is quite close to that for the galaxy M81 and slightly higher than that for the M81 outer disk on average.  The same $R23$-KK04 prescription yields the average oxygen abundance for M81 $12+\log(O/H) = 8.94 \pm 0.09$ ($\sim 1.78\,Z_{\bigodot}$) in \citet{2010ApJS..190..233M}; for \HII~regions in M81 at large radii ($R \geq 0.6\,R_{25}$) $8.86 \pm 0.04$ ($\sim 1.48\,Z_{\bigodot}$) in \citet{2010ApJS..190..233M} and $8.66 \pm 0.34$ ($\sim 0.93\,Z_{\bigodot}$) in \citet{2012MNRAS.422..401P}.  The coincidence of Holmberg\,IX with M81 in metallicity supports the tidal origin of Holmberg\,IX that inherits the metallicity of its parent galaxy M81 from which it was ripped apart.  However, it is puzzling that Holmberg\,IX contains so little (or even no) dust, inconsistent with its high metallicity.  Even if the IR observations were possibly not deep enough to capture effective signals, the dust content for Holmberg\,IX evidently lies far below the normal level for galaxies at this metallicity.  The dustless albeit metal-rich property undoubtedly makes Holmberg\,IX an extraordinary galaxy in the local and distant Universe, and deserves to be explored in further depth.  According to this result,  we consider emissions from Holmberg\,IX unobscured by internal dust, and thus, no correction for dust attenuation is performed to the photometric fluxes in this work.

\subsection{Uniformly Young Stellar Population Demonstrated by UV Diagnosis}\label{Sec_Result_UV}

Stellar population age for Holmberg\,IX is diagnosed by analyzing the four UV images, \emph{GALEX} FUV, NUV, and \emph{Swift}-UVOT UVW2, UVW1, on a spatially resolved basis with the aid of stellar population synthesis modeling.  The UV bands have special advantage in prospecting for young stellar populations, in contrast to previous optical and NIR studies \citep{2008ApJ...676L.113S, 2008ApJ...689..160W, 2015ApJ...809L...1O}.  The following sections present the UV diagnosis in both pictorial and graphic manners.

\subsubsection{2D Color and Age Maps}\label{Sec_Result_UV_2D}

At the first part of the diagnosis, the UV images were smoothed with a boxcar window of $5 \times 5$ pixels (corresponding to $7\arcsec.5 \times 7\arcsec.5$) to reduce noise, and then processed to create 2D maps for color index and stellar population age.  The stellar population age was estimated by fitting $\mathrm{FUV}-\mathrm{NUV}$ and $\mathrm{UVW2}-\mathrm{UVW1}$ with the stellar population synthesis model.  Histograms corresponding to their own 2D maps are charted for statistics of pixels inside the elliptical aperture. The 2D maps for $\mathrm{FUV}-\mathrm{NUV}$ and stellar population age as well as their histograms are shown in Figure \ref{Fig_UVcolor}.

We can see in both of the maps that the galaxy appears with no spatial gradient, ranging 0.15--0.4\,mag in $\mathrm{FUV}-\mathrm{NUV}$ and 80--210\,Myr in age.  Each of the histograms exhibits a single-peak distribution, which is well fitted by a Gaussian profile with the mean $\sim 0.28$\,mag in $\mathrm{FUV}-\mathrm{NUV}$ and $\sim 130$\,Myr in stellar population age, respectively.  The standard deviation of the Gaussian function is $\sim 0.06$\,mag in $\mathrm{FUV}-\mathrm{NUV}$ comparable with the typical uncertainty $\sim 0.08$\,mag induced by background fluctuation, and $\sim 0.10$ in $\log$[Age (yr)] corresponding to $\pm 30$\,Myr at 130\,Myr.  The Gaussian profiles with narrow widths suggest that the spatial variations in the 2D maps are ascribable more to multiple uncertain factors in the SF process or/and the measurements; the stellar population age is therefore considered approximately constant in the galaxy.  From a statistical viewpoint, 130\,Myr is the most probable age for Holmberg\,IX, with the $2 \sigma$ range 80--210\,Myr.

\subsubsection{Color-Color Diagram}\label{Sec_Result_UV_Diag}

The second part of the diagnosis is a graphic analysis of $\mathrm{UVW2}-\mathrm{UVW1}$ as a function of $\mathrm{FUV}-\mathrm{NUV}$ for the subregions (i.e., FUV clusters and diffused regions, described in Section \ref{Sec_Data_Phot}) of Holmberg\,IX.  Figure \ref{Fig_aperUV-UV} shows the photometric apertures in the left panel and the color-color diagram in the right panel, for the FUV clusters, the diffuse regions, the main body of the galaxy as a whole, and the H$\alpha$-emission area in the galaxy; an age sequence reproduced with the GALAXEV model is superimposed for comparison.  Although internal dust attenuation is not taken into account, effects of dust attenuation at the certain amount $A_\mathrm{V}$ = 0.25\,mag with four typical attenuation laws are illustrated for reference.\footnote{In terminological sense, "\emph{attenuation}" is not equivalent to "\emph{extinction}".  People are faced with dust "\emph{attenuation}" when studying extended sources where dust is mixed with stars, while dust "\emph{extinction}" laws are legislated for the Milky Way and the Magellanic Clouds by studying point sources with dust in foreground.  In this Letter, we unify the terminologies into "\emph{attenuation}" even though we actually quote the "\emph{extinction}" laws.  The validity of "\emph{extinction}" laws in characterizing "\emph{attenuation}" is addressed in \citet{2014ApJ...789...76M}.}

From the color-color diagram, we can see that the FUV clusters and the diffuse regions inside the elliptical aperture possess the same regime following the model curve well;\footnote{There are two clusters with $\mathrm{FUV}-\mathrm{NUV} > 0.5$ behaving as the exception. They are separately located on the galactic periphery and suspected to be background sources.} the main body of the galaxy as a whole and the H$\alpha$-emission area are positioned identically at $\mathrm{FUV}-\mathrm{NUV} = 0.3$ mag and $\mathrm{UVW2}-\mathrm{UVW1} = 0.15$\,mag, corresponding to the age $\sim 130$\,Myr by reference to the model, as an average of the subregions.  The overlaps between the different types of regions in this diagram imply all of them indistinguishable in stellar population.  Therefore, the FUV clusters are not definitely younger than the diffuse regions but merely more in the number of stars, so is the H$\alpha$-emission area.  In this situation, all of stars in Holmberg\,IX are supposed to be the same population born from a single SF event $\sim 130$\,Myr ago, as also depicted by the single-peak Gaussian profiles in Figure \ref{Fig_UVcolor}.  The recent occurrence of the global SF coincides with the hypothesis of the tidal origin for this galaxy.  Combining with the above 2D map, we can demarcate for the galactic mainbody the lower age limit $\sim 80$\,Myr that is newly discovered through the UV diagnosis in this work, and the upper limit $\sim 210$\,Myr that is coincident with the previous optical-NIR studies \citep[$< 250$\,Myr,][]{2008ApJ...676L.113S, 2008ApJ...689..160W, 2015ApJ...809L...1O}.  The lack of younger stellar populations suggests a very short duration for the past SF event.  The model curve in Figure \ref{Fig_aperUV-UV} is reproduced with the SFH of an instantaneous SF burst.  Justification for the applied SFH is presented in Appendix \ref{Sec_App} with more complex SFHs compared and discussed.

\subsection{Sparse Ionized Gas Revealed by H$\alpha$ Mapping}\label{Sec_Result_Ha}

For the purpose of exploring ionized gas linking current SF activities, we observed Holmberg\,IX through H$\alpha$ channel in both of the direct narrowband imaging and FPI scanning modes (described in Section \ref{Sec_Data_Ha}).  From a general view provided by the direct narrowband image, the H$\alpha$ emission emerges dim, diffuse, and dispersive all over the galaxy.  In the deep-exposure FPI map, it is spatially resolved into pieces of scattered spots and filaments located at northwest and southeast peripheries; most of the galactic body is left blank.  The H$\alpha$ appearance is opposite to the conspicuous and concentrative UV emission exhibited in the \emph{GALEX} and \emph{Swift}-UVOT images.  Similar H$\alpha$-UV spatial contrast has been found in some other nearby dwarf galaxies and interpreted with SF propagation from galactic centers to outskirts, so that, after a certain timescale, current SF with strong H$\alpha$ arrives in rims surrounding recently formed stellar populations bright in UV \citep{2018MNRAS.478.3386E, 2024MNRAS.529.1138G, 2024MNRAS.529.4930Y}.  Notwithstanding, in this work, such a behavior is observed in a TDG (candidate) for the first time.

H$\alpha$ flux for Holmberg\,IX was extracted via the aperture photometry in the narrowband image and the selective photometry in the FPI image, respectively (described in Section \ref{Sec_Data_Phot}), and the results are listed in Table \ref{Tab_Stat}.  The effective H$\alpha$ area (i.e., detectable in the FPI image), accounting for $\sim 22\%$ of the area enclosed by the elliptical aperture, contributes $\sim 90\%$ of H$\alpha$ but only $\sim 30\%$ of UV emission to the same aperture-enclosed area of the galaxy.

\begin{deluxetable*}{cccccccc}
\tabletypesize{\footnotesize}
\tablecaption{\textbf{Parameters Relevant to SF for Holmberg\,IX}}
\tablewidth{0pc}
\tablehead{
\colhead{Photometry} & \colhead{$M_\star$} & \colhead{$\langle t_\mathrm{\,Age} \rangle$} & \colhead{SFR(H$\alpha$)} & \colhead{SFR(FUV)} & \colhead{$\langle \mathrm{SFR} \rangle_\mathrm{past}$} & \colhead{$b_\mathrm{\,H\alpha}$} & \colhead{$b_\mathrm{\,FUV}$} \\
\colhead{Type} & \colhead{[10$^6\,M_\odot$]} & \colhead{[Myr]} & \colhead{[$10^{-4}\,M_\odot\,yr^{-1}$]} & \colhead{[$10^{-3}\,M_\odot\,yr^{-1}$]} & \colhead{[$M_\odot\,yr^{-1}$]} & \colhead{} & \colhead{} \\
\colhead{(1)} & \colhead{(2)} & \colhead{(3)} & \colhead{(4)} & \colhead{(5)} & \colhead{(6)} & \colhead{(7)} & \colhead{(8)}
}
\startdata
 Aperture & 4.0 & 128 & 5.10~$\pm$~0.30 & 4.80~$\pm$~0.22 & 0.031 & 0.016~$\pm$~0.001 & 0.155~$\pm$~0.007 \\
 Selective & 1.2 & 128 & 4.60~$\pm$~0.14 & 1.47~$\pm$~0.07 & 0.009 & 0.051~$\pm$~0.002 & 0.163~$\pm$~0.008
% Aperture & 4.0 & 127.8 & 5.10 & 4.80 & 0.031 (?mean age?) & 0.016 (?mean age?) & 0.155 (?mean age?) \\
% Selective & 1.2 & 127.8 & 4.60 & 1.47 & 0.009 (?mean age?) & 0.051 (?mean age?) & 0.163 (?mean age?)
% Aperture & 4.0 & 127.8 & 5.10 & 4.80 & 0.020 (?max age?) & 0.026 (?max age?) & 0.240 (?max age?) \\
% Selective & 1.2 & 127.8 & 4.60 & 1.47 & 0.006 (?max age?) & 0.077 (?max age?) & 0.245 (?max age?)
 \enddata
\tablecomments{~
Columns: (1) The types of photometry described in Section \ref{Sec_Data_Phot}; (2) Stellar mass; (3) Stellar population age derived from Figure \ref{Fig_UVcolor} in Section \ref{Sec_Result_UV_Diag}; (4) star formation rate derived from the H$\alpha$ luminosity in Section \ref{Sec_Disc}; (5) star formation rate derived from the FUV  luminosity in Section \ref{Sec_Disc}; (6) Averaged SF rate within past 128 Myr, described in Section \ref{Sec_Disc}; (7) The birthrate parameter defined as $b_\mathrm{\,H\alpha} \equiv \mathrm{SFR(H}\alpha)/\langle\mathrm{SFR}\rangle$, described in Section \ref{Sec_Disc}; (8) The birthrate parameter defined as $b_\mathrm{\,FUV} \equiv \mathrm{SFR(FUV)}/\langle\mathrm{SFR}\rangle$, described in Section \ref{Sec_Disc}.
}
\label{Tab_SF}
\end{deluxetable*}

\section{\textbf{IMPLICATION}}\label{Sec_Implic}

Under average conditions, weak H$\alpha$ emission indicates a currently quiescent SF state.  For Holmberg\,IX, this feature in combination with the UV diagnosis in Section \ref{Sec_Result_UV} potentially manifests global SF already ceased in most areas of the galaxy $\sim 80$\,Myr ago.  In this section, we assess current SF activities reflected by the marginal H$\alpha$ plus the luxuriant FUV emissions, and deduce possible evolutionary pathways for this galaxy.

\subsection{Current Star Formation State}\label{Sec_Implic_SFR}

The differences between the H$\alpha$ and UV emissions in spatial position and radiative intensity inevitably cause considerable discrepancy between SFRs estimated from H$\alpha$ and FUV integrated luminosities.  By taking the aperture photometry and adopting the calibrations in K98, we obtain SFR(H$\alpha$) = ($5.10 \pm 0.30$) $\times 10^{-4}$ $M_\odot\,yr^{-1}$ (corresponding to ($1.09 \pm 0.06$) $\times 10^{-4}$ $M_\odot\,yr^{-1}\,kpc^{-2}$) and SFR(FUV) = ($4.80 \pm 0.22$) $\times 10^{-3}$ $M_\odot\,yr^{-1}$ (corresponding to ($1.03 \pm 0.05$) $\times 10^{-3}$ $M_\odot\,yr^{-1}\,kpc^{-2}$).  Via specifically sampling the H$\alpha$-emission area, the selective photometry plus the K98 calibrations yields SFR(H$\alpha$) = ($4.60 \pm 0.14$) $\times 10^{-4}$ $M_\odot\,yr^{-1}$ (corresponding to ($4.41 \pm 0.13 $) $\times 10^{-4}$ $M_\odot\,yr^{-1}\,kpc^{-2}$), SFR(FUV) = ($1.47 \pm 0.07$) $\times 10^{-3}$ $M_\odot\,yr^{-1}$ (corresponding to ($1.41 \pm 0.06$) $\times 10^{-3}$ $M_\odot\,yr^{-1}\,kpc^{-2}$), which narrows the gap between SFR(H$\alpha$) and SFR(FUV) effectively but remains insufficient.  This examination demonstrates that the spatial mismatch between H$\alpha$ and FUV emissions contributes a major part of the common discrepancy between SFR(H$\alpha$) and SFR(FUV) if galaxies are measured on an integrated basis.  In this instance, different sources with H$\alpha$ and FUV are actually sampled in practice.  The selective photometry enables approaching the same H$\alpha$ and FUV positions but still fails to totally fill the gap between the SFRs.

Different SF timescales traced by H$\alpha$ and FUV luminosities are considered primarily responsible for SFR(H$\alpha$) lower than SFR(FUV), so-called "H$\alpha$-to-FUV deficit", no matter whether the two emissions are spatially displaced or identically located.  From this perspective, SFR(H$\alpha$) takes an average of the latest $\lesssim 10$\,Myr while SFR(FUV) averages past $\lesssim 100$\,Myr (K98).  In this case, SFR(FUV) higher than SFR(H$\alpha$) is a plausible consequence of different SF states tens of Myr ago (active) and at present (quiescent).

%  propagation of SF or quenching?

In order to quantify SF activity levels, we adopt the birthrate parameter $b$ defined as the ratio of current to mean past SFRs \citep[ $b \, \equiv \, \mathrm{SFR}_\mathrm{current}/\langle\mathrm{SFR}\rangle_\mathrm{past}$;][]{1994ApJ...435...22K}.  The proportion valued by $b$ weighs the current star-formation level on the whole lifetime scale of a galaxy and reflects the evolution of SF activities.  In this work, we set the mean past SFR $\langle\mathrm{SFR}\rangle_\mathrm{past}$ equal to $M_\star / t_\mathrm{Age}$, where $M_\star$ is total stellar mass, and $t_\mathrm{Age}$ is the stellar population age.\footnote{The standard definition of the birthrate parameter contains a fraction of stellar mass returned to the galaxy in a stellar generation \citep{1994ApJ...435...22K}.  In this work, we assume a zero return fraction of stellar mass.  Therefore, all mass is newly generated in every SF event, so that the birthrate parameter $b$ reaches the maximum.}  $M_\star$ and $t_\mathrm{Age}$ are obtained with the stellar population synthesis model as best-fit parameters.  The numerator SFR$_\mathrm{current}$ in $b$ is assigned as SFR(H$\alpha$) and SFR(FUV), respectively, which consequently brings on the birthrate parameter representing the SF state in recent $\lesssim 10$\,Myr by $b_\mathrm{\,H\alpha} (\, \equiv \, \mathrm{SFR(H\alpha)}/\langle\mathrm{SFR}\rangle_\mathrm{past})$ and that in recent $\lesssim 100$\,Myr by $b_\mathrm{\,FUV} (\, \equiv \, \mathrm{SFR(FUV)}/\langle\mathrm{SFR}\rangle_\mathrm{past})$.

The total stellar mass, stellar population age, SFR(H$\alpha$), SFR(FUV), $\langle\mathrm{SFR}\rangle_\mathrm{past}$, $b_\mathrm{\,H\alpha}$, and $b_\mathrm{\,FUV}$ are derived with the two types of measurements (the aperture photometry and the selective photometry) and listed in Table \ref{Tab_SF}.  We can see $b_\mathrm{\,FUV}$ higher than $b_\mathrm{\,H\alpha}$ by $\sim 10$ times with the aperture photometry and $\sim 3$ times with the selective photometry; in either case, however, it is less than 0.3 and indicates that the galaxy has lain in a quiescent state not only in the past 10\,Myr but also since tens of Myr ago \citep[$b < 0.3$ for quiescent galaxies;][]{2004MNRAS.349..769K}.

\subsection{Subsequent Evolution and Ultimate Fate}\label{Sec_Implic_Evol}

The global SF dearth albeit the overall infancy for stellar populations requires rapid SF quenching, for which the reason seems to be a conundrum, particularly in view that the galaxy resides in a gas-rich environment \citep[$M_\mathrm{HI} = 3.28 \times 10^{8} M_\odot$ and $M_\mathrm{HI}/M_\star > 40$ for Holmberg\,IX, presented in][]{2008ApJ...676L.113S}.  At the same time, the currently quiescent state conflicts with the commonly accepted requisite of continuous SF for TDGs \citep[][and references therein]{1999IAUS..186...61D} and raises a question on subsequent evolution and ultimate fate of Holmberg\,IX.  Without SF continuing, Holmberg\,IX is supposed to stay in a transient status at present and be transforming into a long-lived dwarf spheroidal galaxy due to gravitational contraction similar to those prevailing in the Local Group,\footnote{In the Local Group, although there is no TDG, spheroidal galaxies are likely to originate from tidal events, which has been predicted by a number of numerical simulations \citep[e.g.,][]{1998ARA&A..36..435M, 2012MNRAS.424.1941C, 2014MNRAS.442.2419Y}.}, or to the opposite, dissolving into an ultra-diffuse galaxy (UDG) or intergalactic stellar stream because of centrifugal effects (and perhaps vanish in the Universe eventually).  At this present moment, we are probably witnessing a rarely captured phenomenon for a TDG.

\section{\textbf{DISCUSSION}}\label{Sec_Disc}

The above assessment and deduction rely on a high-mass SF scenario, which ascribes weak H$\alpha$ but strong FUV emanations (or H$\alpha$-to-FUV deficit) to SFR low in recent $\lesssim 10$\,Myr but high tens of Myr ago with bursty SF.  In this case, aging of stellar populations leads to lack of high-mass, short-lifetime ionizing stars, which weakens H$\alpha$ but sustains longer-time luminescence of FUV \citep{2000MNRAS.312..442S, 2001ApJ...548..681B, 2004A&A...421..887I, 2004MNRAS.350...21S, 2012ApJ...744...44W}.  However, some observations of dwarf galaxies by resolving individual stars have suggested a more complicated situation that high-mass ionizing stars are not always absent for the H$\alpha$-to-FUV deficit \citep{2008ApJ...689..160W, 2009ApJ...695..561M, 2010ApJ...721..297M, 2010ApJ...724...49M}.  In practical studies, IMF is usually assumed in a universal form with massive SF dominating, but if it turns out to be a variant in some circumstance, the speculation about galactic properties would be different.  Besides IMF forms, a variety of other factors including SF modalities, interstellar medium conditions, observational bias, etc. are also able to interpret the H$\alpha$-to-FUV deficit recognized in a number of dwarf galaxies and extended outer disks (so-called extended UV disks, or XUV disks) of spiral galaxies \citep[e.g.,][and references cited in this section]{2024ApJS..271....2H}.

Instead of currently quiescent SF activities, the alternative explanations of H$\alpha$-to-FUV deficit allow the current occurrence of intense SF in agreement with the common expectation for TDGs.  For unity of SFR(H$\alpha$) and SFR(FUV), \citet[][hereafter denoted as L09]{2009ApJ...706..599L} have recalibrated $L(\mathrm{H}\alpha)$ to SFR for dwarf galaxies with $L(\mathrm{H}\alpha) < 2.5 \times 10^{39}$\,erg\,s$^{-1}$.  By applying the L09 calibration to Holmberg\,IX in our work, we obtain SFR(H$\alpha$) = ($3.08 \pm 0.53$) $\times 10^{-3}$ $M_\odot\,yr^{-1}$ (corresponding to ($6.57 \pm 1.13$) $\times 10^{-4}$ $M_\odot\,yr^{-1}\,kpc^{-2}$) for the aperture photometry, indeed reducing the offset from SFR(FUV), and SFR(H$\alpha$) = ($2.89 \pm 0.33$) $\times 10^{-3}$ $M_\odot\,yr^{-1}$ (corresponding to ($2.77 \pm 0.32$) $\times 10^{-3}$ $M_\odot\,yr^{-1}\,kpc^{-2}$) for the selective photometry, even exceeding SFR(FUV).  Nevertheless, applicability of the L09 prescription to all dwarf galaxies is uncertain, and the resultant current SF activity for Holmberg\,IX is questionable until solid evidence is acquired.  In below subsections, we elucidate all the other attributions of the H$\alpha$-to-FUV deficit and discuss the suitability for Holmberg\,IX.

\subsection{Low-mass or/and Stochastic SF}

A low H$\alpha$ amount is most directly ascribed to inherent shortage of high-mass ionizing stars.  In addition to aging of stellar populations formed on a short or instantaneous timescale as suggested above, continuous SF with IMF in a low-mass dominant form is another mechanism innately preventing massive SF and consequently inducing H$\alpha$-to-FUV deficit.  The variant IMF has been not only reproduced by stellar population synthesis modeling \citep{2003ApJ...598.1076K, 2005ApJ...625..754W, 2006MNRAS.365.1333W, 2007ApJ...671.1550P, 2009MNRAS.395..394P} but also observed in local galaxies with low surface brightness \citep{2008ApJ...675..163H, 2009ApJ...695..765M} and galactic XUV disks \citep{2020MNRAS.491.2366B, 2024A&A...681A..76R}; still, it is not regarded universally effective \citep{2010AJ....139..447H}.

Even if stars keep forming with invariantly normal IMF, a lack of high-mass stars is also likely caused by stochasticity in SF events \citep{2012ApJ...745..145D, 2014MNRAS.444.3275D}.  It is possible for random sampling of constant IMF at low SFR to undersample the number of high-mass ionizing stars (L09, and references therein), but the stochastic SF seems, still, not to suffice for the explanation either \citep{2012ApJ...744...44W}.  Therefore, two or more combinative effects of bursty, low-mass, or/and stochastic SF are considered more reasonable for restricting massive SF and more realistic for explaining observational H$\alpha$-to-FUV deficit \citep{2011ApJ...741L..26F, 2014MNRAS.444.3275D, 2016ApJ...817..177L}.  In this case, Holmberg\,IX is likely to keep forming stars for maintaining long-time survival, which hints there are possibly TDGs of the same type that failed to be recognized in the Universe.

\subsection{Loss of Ionizing Photons}

If high-mass ionizing stars are always sufficient, H$\alpha$-to-FUV deficit is likely caused by loss of ionizing photons.  The most probable mechanism is leakage of Lyman continuum in thin, porous, or/and (semi)transparent interstellar medium, which tends to decrease SFR(H$\alpha$) by up to 25\% at fixed SFR(FUV) \citep{2012MNRAS.423.2933R}.  This mechanism has been applied to dwarf galaxies and XUV disks as an alternative explanation of H$\alpha$-to-FUV deficit \citep{2010AJ....139..447H, 2011AJ....142..121H, 2013AJ....146...92H}.  However, it seems not possible for Holmberg\,IX located in a gas-rich environment where interstellar medium is rigorous and tight.  Even though the HII regions are leaky and ionizing photons escape from natal areas, most of these photons still stay within the galaxy and supply diffuse emission lines, which is accounted for in integrated measurements.

Dust attenuation before gas ionization is another possible reason for the ionizing-photon loss and the decrease in SFR(H$\alpha$) \citep[by $\sim 10$--$15\,\%$][]{2004A&A...421..887I}.  However, since dust content in Holmberg\,IX is unobservable (as presented in Section \ref{Sec_Result_IR}), this circumstance seems also unrealistic for this galaxy as well.

\subsection{Underestimates of Existing H$\alpha$ Emission}

In addition to lack of ionizing stars and loss of ionizing photons, underestimates of H$\alpha$ emission potentially contribute to the offset between observed and expected H$\alpha$ levels, particularly for low surface-brightness systems \citep{2010AJ....139..447H}.  However, much deeper imaging of dwarf galaxies than previous surveys has evinced that the unaccounted part of H$\alpha$ emission (if any) compensates for only 5\% flux at maximum and thus fails to suffice for a self-contained explanation \citep{2016ApJ...817..177L}.  Nay, our ultra-deep FPI observations \citep[with the sensitivity limit lower than that in][]{2016ApJ...817..177L} and the net emission-line extraction from the 3D datacube guarantee the completeness of sampling H$\alpha$ emission and thereof prevent the underestimation.

\section{\textbf{SUMMARY}}\label{Sec_Sum}

Via new observations deeply mapping H$\alpha$ emission in Holmberg\,IX and combinatory analysis of archival/published data, we reveal a series of unique characteristics for this TDG candidate, distinct from other TDGs ever known.  In the results, the galaxy appears extremely dustless, disproportionate to its rich gas and metal, with globally young stellar populations but currently inactive SF reflected by H$\alpha$-to-FUV deficit.  In the galactic mainbody, all stars are suggested to be born from a bursty SF event $\sim 130$\,Myr ago, which then rapidly ceased, demarcating the lower age limit $\sim 80$\,Myr; younger stars reside in outskirts where mild SF remains as an aftermath of SF radial propagation.  The recent quenching seems like a conundrum in view of the gas-rich location of the galaxy.

On a normal massive-SF assumption, the present quiescence conflicts with the common expectation for TDGs hosting continuous SF, which hints Holmberg\,IX currently staying in a rarely captured stage transforming into a dwarf spheroidal galaxy or oppositely into a(n) (ultra-)diffuse (intergalactic) system, which will probably dissolve in the end.  Alternatively, by assuming a peculiar form of IMF, it is also possible for Holmberg\,IX to keep forming low-mass stars with H$\alpha$-to-FUV deficit and hence be able to maintain the current status for long-time survival.  In whichever situation in reality, differing from conventional understandings, Holmberg\,IX deserves further investigations of its past experience, present state, and future destiny, as well as explorations of analogous TDGs in the Universe.

\acknowledgments

We are grateful to the anonymous reviewer for carefully reading the manuscript and offering very helpful comments, which have led to a necessary improvement in this Letter.  This work is supported by the National Natural Science Foundation of China (NSFC, Nos. 11603007, U2031106, 12233001) and the China Manned Space Program (CMS-CSST-2025-A09).  Observations with the SAO RAS telescopes are supported by the Ministry of Science and Higher Education of the Russian Federation. The renovation of telescope equipment is currently provided within the national project "Science and Universities". The work on the 6 m telescope data reduction and analysis was performed as part of the SAO RAS government contract approved by the Ministry of Science and Higher Education of the Russian Federation.  O.E. acknowledges funding from the Deutsche Forschungsgemeinschaft (DFG, German Research Foundation) -- project ID 541068876.  Y.-W.M. specially thanks Ying-Jie Peng, Wei Zhang for scientific discussion, and Wei Zhang, Qi-Ming Wu for technical assistance.  This research has made use of the NASA/IPAC Extragalactic Database (NED), which is operated by the Jet Propulsion Laboratory, California Institute of Technology, under contract with the National Aeronautics and Space Administration. This research has also made use of NASA's Astrophysics Data System.

\vspace*{1.15cm}
\appendix

\section{\textbf{JUSTIFICATION FOR THE ASSUMED STAR FORMATION HISTORY}}\label{Sec_App}

In Section \ref{Sec_Result_UV}, we adopt the stellar population synthesis modeling with an instantaneous SF burst to characterize the spatially resolved UV colors.  The statistical analysis on a pixel-by-pixel basis suggests the galaxy born from a single bursty SF event $\sim 130$\,Myr ago most probably.  Regions on a subgalactic scale are intrinsically simple in stellar population and SFH, which enables the assumption of an instantaneous SF burst naturally suitable for the galactic subregions in this work.  Notwithstanding, it is still necessary to inspect whether it is possible for the subregions to experience more complex SFHs and whether there is an artifact caused by the assumed simple SFH.  The inspection is presented in this appendix.

\begin{figure*}[!ht]
\centering
\hspace*{-1.75cm}
\includegraphics[width=0.4\columnwidth]{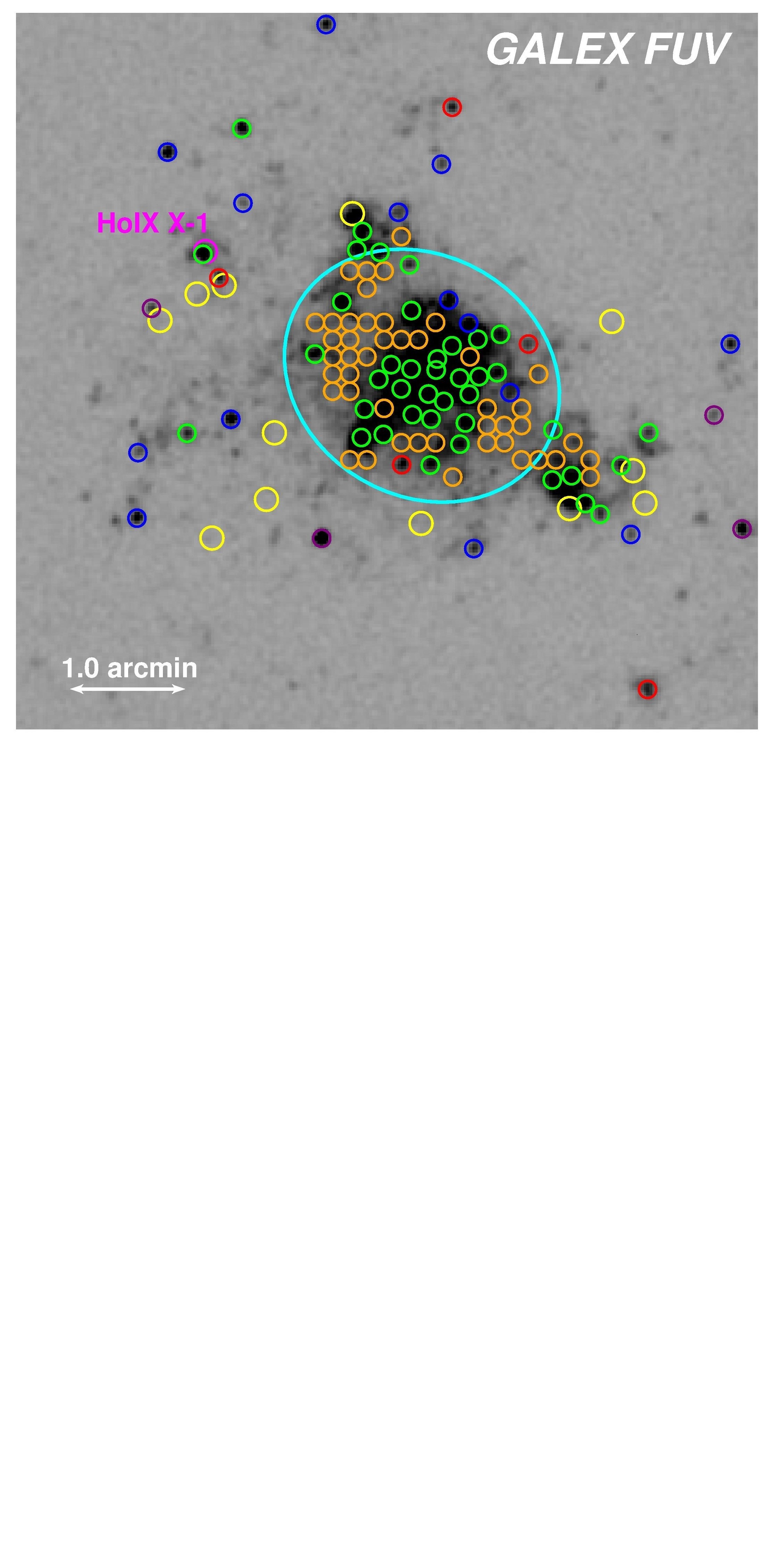}
\hspace*{-1.0cm}
\includegraphics[width=0.7\columnwidth]{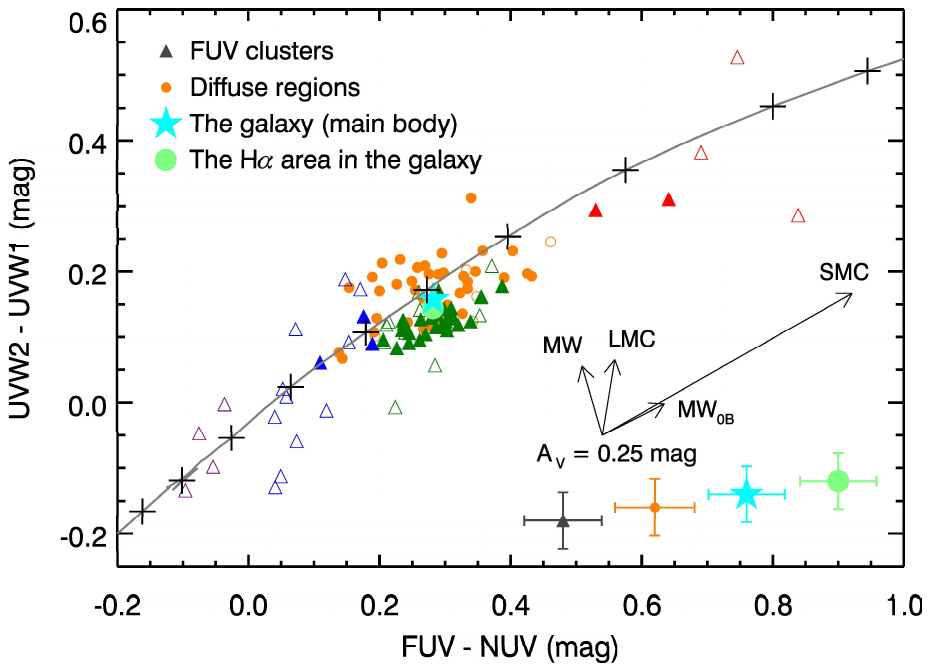}
\hspace*{-2.75cm}
\\[-7.25cm]
\caption{Left: \emph{GALEX}-FUV inverse-brightness image for Holmberg\,IX, where the FUV clusters, the diffuse regions, and the main body of the galaxy as a whole, the H$\alpha$-emission area in the galaxy, HoIX\,X-1, and the foreground stars are marked. Cyan ellipse, yellow and magenta circles are the same as assigned in Figure \ref{Fig_UVHaIR}; All of the other circles represent photometric apertures for the FUV clusters and the diffuse regions. The apertures for the FUV clusters are color-coded by $\mathrm{FUV}-\mathrm{NUV}$: $\mathrm{FUV}-\mathrm{NUV} < 0.0$\,mag (purple), $0.0 \leq \mathrm{FUV}-\mathrm{NUV} < 0.2$\,mag (blue), $0.2 \leq \mathrm{FUV}-\mathrm{NUV} < 0.4$\,mag (green), and $\mathrm{FUV}-\mathrm{NUV} \geq 0.4$\,mag (red); the apertures for the diffuse regions are in orange. The scale ruler at the bottom-left corner indicates the 1$\arcmin$.0 length; north is up, and east is to the left.
Right: $\mathrm{UVW2}-\mathrm{UVW1}$ as a function of $\mathrm{FUV}-\mathrm{NUV}$ for the FUV clusters (triangles), the diffuse regions (orange circles), the main body of the galaxy as a whole (cyan-filled star), and the H$\alpha$-emission area in the galaxy (green-filled circle), with a GALAXEV model curve superimposed (gray-solid line). The FUV clusters and the diffuse regions inside the cyan-elliptical aperture are represented by filled symbols, while those outside that are represented by open ones. The color-coding for the FUV clusters is the same as in the left panel. The model curve shows an age sequence with black crosses marking the positions of the ten ages, 3.8, 10, 20, 40, 80, 130, 210, 300, 360, and 400\,Myr from left to right.  The arrows show the vectors predicting the effects of the dust attenuation at the amount $A_\mathrm{V}$ = 0.25\,mag for the Milky Way (MW), the Large Magellanic Cloud (LMC), and the Small Magellanic Cloud (SMC) laws, as well as the attenuation law with the MW-type slope but no 2175\,$\mathrm{\AA}$ bump (denoted as MW$_\mathrm{0B}$ in the diagram). The quotation and the construction of the attenuation laws are presented in \citet{2014ApJ...789...76M}. The median errors for the FUV clusters (blue), the diffuse regions (orange), the main body of the galaxy as a whole (cyan), and the H$\alpha$-emission area in the galaxy (green) are plotted at the bottom-right corner.} \label{Fig_aperUV-UV}
\end{figure*}

\begin{figure*}[!ht]
\centering
\includegraphics[width=\columnwidth]{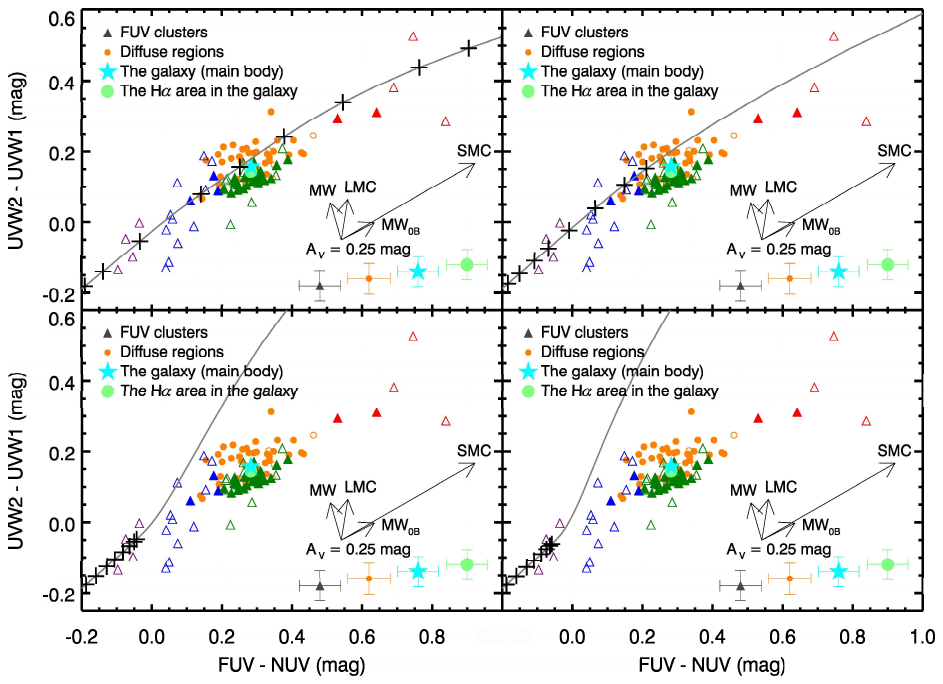}
\\[-10.25cm]
\caption{$\mathrm{UVW2}-\mathrm{UVW1}$ as a function of $\mathrm{FUV}-\mathrm{NUV}$.  The symbols, the color-coding of the data points, and the arrows are the same as assigned in the right panel of Figure \ref{Fig_aperUV-UV}, but the superimposed GALAXEV model curve is reproduced with exponentially decreasing SFR with the SF timescale $\tau_\mathrm{SF}$ = 0.01 (top-left
panel), 0.1 (top-right panel), 0.5 (bottom-left panel), and 1\,Gyr (bottom-right panel).  The black crosses mark the positions of the same ages as assigned in the right panel of Figure \ref{Fig_aperUV-UV}, with the rightmost cross representing 400\,Myr, but some of the left-side crosses in Figure \ref{Fig_aperUV-UV} exceed the axis range in this figure.} \label{Fig_UV-UV-SFH}
\end{figure*}

Figure \ref{Fig_UV-UV-SFH} shows the same color-color diagrams as plotted in the right panel of Figure \ref{Fig_aperUV-UV}, but superimposed is the GALAXEV model curve reproduced by exponentially decreasing SFR with the four types of SF timescale $\tau_\mathrm{SF}$ = 0.01, 0.1, 0.5, and 1\,Gyr, respectively.  Combining Figures \ref{Fig_aperUV-UV} and \ref{Fig_UV-UV-SFH}, it is apparent to see that simpler SFHs fit the data locus better, where the instantaneous SF burst provides the best fit.  The data points for the integrated measurements are located the closest to the model track of the instantaneous SF burst and overlap the marked 130\,Myr model point, which validates the bursty SFH and the age for the galaxy as a whole derived from the statistics of the galactic subregions.

On the other hand, among all forms of SFHs, the instantaneous SF burst yields the reddest colors for stellar populations at a constant age; any more complex SFHs lead to older ages for fixed color indices \citep[detailed analyses of SFH impacts on UV properties for stellar populations are presented in Section 6 in][]{2012ApJ...757...52M}.  According to the previously derived age $< 250$\,Myr for Holmberg\,IX \citep{2008ApJ...676L.113S, 2008ApJ...689..160W, 2015ApJ...809L...1O}, we are also able to justify the conclusion about the SFH and the age drawn in this work.  If Holmberg\,IX and its subregions experience more complex SFHs, then the given UV colors require the galaxy in possession of stars older than 250\,Myr, which should have been detected through the optical-NIR channels.

\clearpage

\clearpage

\clearpage

\clearpage

\end{document}